\documentclass[pre,twocolumn,showpacs,showkeys,amsmath,amssymb]{revtex4}
\usepackage[english]{babel}
\usepackage{graphicx}
\newcommand{\onefig}[1]{\centering{\includegraphics[width=0.96\columnwidth]{#1}}}

\newcommand{\partialt}[1]{\partial_t #1}
\newcommand{\partialx}[1]{\partial_x #1}
\newcommand{\partialxx}[1]{\partial^2_x #1}

\newcommand{\subcrit}{\mathrm{c}}
\newcommand{\sublinmax}{\mathrm{lin}}
\newcommand{\subkapyor}{\mathrm{KY}}
\newcommand{\subS}{\mathrm{s}}
\newcommand{\subU}{\mathrm{u}}
\newcommand{\supT}{\mathrm{T}}

\newcommand{\myexp}[1]{\mathrm{e}^{#1}}
\newcommand{\myii}{\mathrm{i}}

\renewcommand{\vec}[1]{#1}
\newcommand{\mtr}[1]{#1}

\begin{document}

\title{Violation of hyperbolicity in a diffusive medium with local
hyperbolic attractor}

\author{Pavel V. Kuptsov}\email[Corresponding author. Electronic address:]{p.kuptsov@rambler.ru}
\affiliation{
  Department of Informatics, Saratov State Law Academy,
  Volskaya 1, Saratov 410056, Russia}
\author{Sergey P. Kuznetsov}
\affiliation{
  Kotel'nikov Institute of Radio Engineering and
  Electronics of RAS, Saratov Branch, Zelenaya 38, Saratov 410019,
  Russia}



\date{\today}

\begin{abstract}
Departing from a system of two non-autonomous amplitude equations,
demonstrating hyperbolic chaotic dynamics, we construct a 1D
medium as ensemble of such local elements introducing spatial
coupling via diffusion. When the length of the medium is small,
all spatial cells oscillate synchronously, reproducing the local
hyperbolic dynamics. This regime is characterized by a single
positive Lyapunov exponent. The hyperbolicity survives when the
system gets larger in length so that the second Lyapunov exponent
passes zero, and the oscillations become inhomogeneous in space.
However, at a point where the third Lyapunov exponent becomes
positive, some bifurcation occurs that results in violation of the
hyperbolicity due to the emergence of one-dimensional
intersections of contracting and expanding tangent subspaces along
trajectories on the attractor. Further growth of the length
results in two-dimensional intersections of expanding and
contracting subspaces that we classify as a stronger type of the
violation. Beyond of the point of the hyperbolicity loss, the
system demonstrates an extensive spatiotemporal chaos typical for
extended chaotic systems: when the length of the system increases
the Kaplan-Yorke dimension, the number of positive Lyapunov
exponents, and the upper estimate for Kolmogorov-Sinai entropy
grow linearly, while the Lyapunov spectrum tends to a limiting
curve.
\end{abstract}

\maketitle

\section*{Introduction}

One of the central concepts in mathematical theory of dynamical
systems relates to hyperbolic strange attractors. Tangent space of
each point of such an attractor splits into expanding and
contracting subspaces, and this splitting is invariant. Dynamics
on a hyperbolic attractor is structurally stable, i.e, is
insensible to variations of parameters. It manifests strong
stochastic properties and allows detailed theoretical
analysis~\cite{KatHass,GuckHolm}.

During the last 40 years hyperbolic attractors were considered
rather as idealized model of perfect chaos. Though some artificial
systems with hyperbolic attractors were known, they were useless
for practical applications because of complicated construction.
Recently, a realistic system was suggested and implemented as
electronic device, dynamics of which in stroboscopic description
is associated with attractor of Smale-Williams
type~\cite{KuzHyp,HypExper}. Attractor of this system is
hyperbolic as proven numerically by verification of
the cone criterion~\cite{KuzSat07}. In
paper~\cite{KupKuzSat} the amplitude equation for this system was
studied, and the hyperbolicity was also proven by the method of
cones. (Some other models based on the same principle are
considered in Refs.~\cite{IsYuJalKuz06,KuzPik07,KuzPik08,KuzPon08}.)

Traditionally, studies of hyperbolic dynamics are mostly
concentrated on low dimensional systems. Many topics concerning
spatiotemporal chaos, though attracted a lot of interest, remain
open~\cite{LSYoung08}. In this paper we address a problem of
survival of hyperbolicity of a spatiotemporal system when the
length of the system grows. We consider a 1D extended system
composed of local elements possessing a hyperbolic attractor that
is based on the amplitude equations from~\cite{KupKuzSat}. The
spatial coupling is introduced via diffusion. In fact, a system we
study is a set of two coupled non-autonomous Ginzburg-Landau
equations of special form.

We are aware of two numerical methods for reliable verification of
hyperbolicity. The first one is the method based on the cone
criterion~\cite{KuzSat07}, which employs directly the rigorous
theorem, and, hence, looks preferable. Unfortunately, the method
is appropriate only for low-dimensional systems, while its
extension to systems of many degrees of freedom seems to be
abundantly sophisticated. The second method is based on a recently
suggested routine of computing of covariant Lyapunov
vectors~\cite{Ginelli07}. These vectors are associated with
Lyapunov exponents and indicate directions of contracting and
expanding manifolds at each point of an attractor. If these
vectors are known, angles between each contracting and each
expanding direction can be computed and the minimal one can be
found. The attractor is interpreted as non-hyperbolic, if the
distribution of these angles does not vanish at the origin. In
fact, this is only a sufficient condition because the converse is
not true. In the present paper we apply more subtle approach based
on computation of so called principal angles~\cite{GolubLoan},
that allows to detect a tangency of two arbitrary vectors from
contracting and expanding subspaces.

The paper is organized as follows. In Sec.~\ref{sSystem} we
introduce the system and briefly discuss its local dynamics. Also
we describe a numerical method applied to find solutions to the
system. Sec.~\ref{sLSA} represents linear stability analysis. The
critical length of the system is determined where a spatially
homogeneous solution becomes unstable with respect to non-uniform
perturbation. Sec.~\ref{sSptm} is devoted to illustrations of
spatiotemporal dynamics. The main part of the paper is
Sec.~\ref{sLyap}, where we develop the Lyapunov analysis. We
discuss distributions of minimal angles between contracting and
expanding tangent subspaces on the attractor. Also, dependencies
of Lyapunov exponents, Kaplan-Yorke dimension and Kolmogorov-Sinai
entropy on the length of the system are considered. In
Sec.~\ref{sRes} we summarize the obtained results and outline
perspectives for further investigations.

\section{\label{sSystem}The model and numerical method}

Let us start with a physical model, demonstrating hyperbolic
dynamics, suggested by Kuznetsov in Ref.~\cite{KuzHyp}. The model
consists of two coupled non-autonomous van der Pol oscillators
that are parametrically influenced by an external periodic force.
The oscillators become active turn by turn, and pass the
excitation each other in such way that the phase of oscillations
is doubled after each period of the forcing. In
Ref.~\cite{KupKuzSat} the amplitude equations for this system are
derived that read
\begin{equation}
  \label{eOdeSystem}
  \begin{gathered}
    \dot{a} =   A \cos (2\pi t/T) a - |a|^2 a - \myii \epsilon \, b, \\
    \dot{b} = - A \cos (2\pi t/T) b - |b|^2 b - \myii \epsilon \, a^2.
  \end{gathered}
\end{equation}
In this paper we study a spatially extended analog of these
equations, supplying them with the second spatial derivatives.

So, we consider two coupled non-autonomous Ginzburg-Landau
equations:
\begin{equation}
  \label{eTheSystem}
  \begin{gathered}
    \partialt{a} =   A \cos (2\pi t/T) a - |a|^2 a - \myii \epsilon \, b + \partialxx{a}, \\
    \partialt{b} = - A \cos (2\pi t/T) b - |b|^2 b - \myii \epsilon \, a^2  + \partialxx{b}.
  \end{gathered}
\end{equation}
Here $a\equiv a(x,t)$ and $b\equiv b(x,t)$ are complex dynamical
variables whose behavior is the subject of interest. Coefficients
at linear terms undergo periodic variation with period $T$ and
amplitude $A$. The parameter modulation takes place in opposite
phase for $a$ and $b$. When the first subsystem is excited, the
second one is relaxed and vice versa. The forcing is supposed to
be slow, i.e., the half period $T/2$ is much longer then a
transient time of the excitation. The second terms in the
right-hand parts of the equations provide saturation of
instabilities in the excited subsystems. Additionally, there are
terms responsible for coupling between the components $a$ and $b$;
the intensity of the coupling is controlled by $\epsilon$. The
coupling is asymmetric, being quadratic from $a$ to $b$ and linear
in the inverse direction. Finally, the last terms in the
right-hand parts introduce diffusion, that is responsible for the
spatial distribution of local oscillations. The diffusion
coefficients of the subsystems are equal to $1$. We study the
system in a limited spatial domain $0 \leq x \leq L$. The boundary
conditions are
\begin{equation}
  \label{eBoundCond}
  \left(\partialx{a}\right)_{|x=0,L}=\left(\partialx{b}\right)_{|x=0,L}=0.
\end{equation}

Let us briefly discuss a local dynamics of the system. Consider
Eqs.~\eqref{eOdeSystem}. (A more detailed study can be found in
Ref.~\cite{KupKuzSat}.) Due to the presence of periodic forcing in
\eqref{eOdeSystem}, it is natural to introduce a stroboscopic map:
we split the continuous time into steps of length $T$, and
consider a sequence of states of the system at the beginnings of
these steps. Define phases within the interval $[0,2\pi)$: $\phi =
\arg a$, $\psi = \arg b$. Suppose at some instant the first
oscillator is excited, and its amplitude $|a|$ is high. Then, the
second one is suppressed, and its amplitude $|b|$ is small. The
coefficients in \eqref{eOdeSystem} are real, except the coupling
term. It means that the phases can vary only as a result of
interaction between subsystems. But, when $a$ is excited, $|b|$ is
small, and its action on $a$ is negligible. Thus, the phase of $a$
remains approximately constant during the excitation stage. On the
contrary, the influence of the excited $a$ on the suppressed $b$
is strong. The coupling term is proportional to $a^2$. It means
that after the half period $T/2$ at the threshold of its own
excitation the oscillator $b$ inherits a doubled phase of $a$
(also the phase gets a shift $-\pi/2$ because of the imaginary
unit at the coupling term). Now the roles of the subsystems are
exchanged. The phase of $b$ remains constant when this subsystem
is excited and at the end, after the other $T/2$, the phase is
returned back to $a$ through a linear coupling term (also with the
shift $-\pi/2$). As a result, the first oscillator $a$ doubles its
phase during the period~$T$. This discussion allows to write down
a map for a series of phases $\phi_n = \arg a(nT)$ that are
measured over the time step $T$:
\begin{equation}
  \label{eBernulMap}
  \phi_{n+1}=2\phi_n - \pi  \mod 2\pi.
\end{equation}
Up to a constant term (that can be eliminated by a shift of the
origin of the phase) this map coincides with the well known
Bernoulli map~\cite{Schuster84,Ott93}. It demonstrates chaotic
dynamics, and the chaos is homogeneous: a rate of exponential
divergence of two close trajectories is identical at each point of
the phase space, being equal to $\ln 2$.

Getting back to the continues system~\eqref{eOdeSystem}, we
estimate its largest Lyapunov exponent as
\begin{equation}
  \label{eHypLyapunov}
  \lambda_0 = \ln 2 / T.
\end{equation}
The described mechanism of phase doubling presumes a hyperbolic
nature of the dynamics of~\eqref{eOdeSystem}. The numerical
verification of the cone criterion, that has been preformed
in~\cite{KupKuzSat}, confirms this.

Before starting an analysis of the system~\eqref{eTheSystem}, let
us discuss a numerical method applied to find its solutions.
Formally, our equations can be classified as parabolic PDE.
Typical recommendation of handbooks for such equations is the
Crank-Nicolson method which is absolutely stable and provides the
second order of local approximation both in space and in time.
This method is semi-implicit, i.e., a solution at a new level
$t_{k+1}$ is expressed via previous solution at $t_k$ as a set of
algebraic equations, so that values from all spatial points on
both levels are involved into this equation set. If PDE is linear,
these equations are linear too. But application of this approach
to non-linear PDEs, like ours, gives rise to a set of non-linear
algebraic equations that requires much more computational efforts.
Usually, one simplifies the problem by neglecting terms, being
non-linear with respect to unknown variables. The resulting
numerical scheme is semi-implicit for linear part of initial PDE
and explicit for non-linear part. Unfortunately, this simplified
``quasi Crank-Nicolson'' method is not absolutely stable.
Sometimes everything goes fine, but sometimes, usually when the
system is far beyond the instability threshold, the solution
diverges. In this paper we do not neglect the non-linearity and
develop a true semi-implicit scheme. At each time step we solve a
set of non-linear equations via the Newton-Raphson iterations. The
seed for the iterations is found from the mentioned simplified
method. The iterations converge very fast. Normally, it takes $2$
or $3$ repetitions to solve the non-linear equations with the
accuracy $10^{-5}$ or even better. The idea of the described
method can be found in books on numerical analysis, e.g.,
\cite{Kalitkin78,Ames77}. Though the method is a bit complicated,
this is compensated by its high accuracy and stability.

Below different characteristic values are calculated as functions
of the length of the system $L$. Varying $L$, we need to choose
some strategy of simultaneous variation of parameters of a
numerical mesh. One way is to keep constant number of points of
the mesh $N$ and compute space step as $\Delta x = L/(N-1)$. The
other way is to fix the step $\Delta x$ and find $N$ for each $L$
as $N = 1 + \lceil L/\Delta x \rceil$, where $\lceil\cdot\rceil$
means ceiling (to get a consistent numerical scheme one also needs
to adjust actual value of $\Delta x$ for the equality $\Delta x
(N-1) = L$ to fulfill). In our simulations we always keep constant
$N$. This strategy seems to be preferable because the number of
degrees of freedom of the numerical model remains constant;
obviously, it is equal to $2N$ (traditionally defined as a half of 
a total order of the set of differential equations). 
So, we can be sure that phenomena,
observed when $L$ is varied, emerge due to a transformation of an
inner structure of attractor, and they can not be attributed to
just an extensive increase of degrees of freedom. The time step
$\Delta t$ can be either constant or attached to $\Delta x$. When
$\Delta t$ is sufficiently small, these two ways produce identical
results. We shall hold the time step at $\Delta t\approx 0.01$.
(Additionally, a small adjustment is also made to fit an integer
number of steps into the observation interval). Though this is
redundantly small value to obtain solutions to the
system~\eqref{eTheSystem}, but this is needed to 
estimate correctly its minor
Lyapunov exponents.

\section{\label{sLSA}Linear stability analysis}

Standard linear stability analysis of autonomous spatially
extended active system requires a consideration of small
perturbations to a homogeneous steady state. Existence of
perturbation modes with positive growth rates indicates the
instability of the homogeneous state. Our system does not have a
steady state, and its dynamics is chaotic in time. Oscillations
can be either homogeneous or irregular in space. Our aim is to
find the conditions for a transition from one regime to another,
utilizing ideas of the standard analysis.

Suppose that the system is infinite in space and its initial state
is uniform. Prepared in this way, the system obviously
demonstrates homogeneous oscillations; at any spatial point the
dynamics can be described by the ODE system~\eqref{eOdeSystem}.
Let us consider an inhomogeneous perturbation to these
oscillations. We need to seek a solution composed as a sum of a
homogeneous part, say, $a_0(t)$ and $b_0(t)$, and a sinusoidal
mode of perturbation with real wave number $k$ and real growth
rate $\sigma(k)$. The system is chaotic; thus, instead of usual
assumption of time periodicity of small perturbation, we introduce
small amplitudes $\tilde{a}(t)$ and $\tilde{b}(t)$ and require
them neither grow nor decay, in average. It means that there exist
two constants, $0<K<M<\infty$, such that $K<|\tilde{a}(t)|<M$ and
$K<|\tilde{b}(t)|<M$ for $t>0$. So, we set:
\begin{equation}
  \label{eLinSol}
  \begin{gathered}
    a(x,t)=a_0(t)+\tilde{a}(t) \myexp{\sigma(k) t -\myii kx}, \\
    b(x,t)=b_0(t)+\tilde{b}(t) \myexp{\sigma(k) t -\myii kx}.
  \end{gathered}
\end{equation}
After substitution~\eqref{eLinSol} to~\eqref{eTheSystem}, we
exclude non-linear terms in $\tilde{a}$ and $\tilde{b}$, supposed
to be small, and obtain a set of linear ODE for complex amplitudes
of perturbation:
\begin{equation}
  \label{eLinModes}
  \begin{gathered}
    \dot{\tilde{a}} = (A \cos (2\pi t/T)-\lambda_0) \tilde{a} -
      2|a_0|^2 \tilde{a} - a_0^2 \tilde{a}^* - \myii \epsilon \tilde{b}, \\
    \dot{\tilde{b}} = (-A \cos (2\pi t/T)-\lambda_0) \tilde{b}  -
      2|b_0|^2 \tilde{b} - b_0^2 \tilde{b}^* - 2\myii \epsilon a_0\tilde{a},
  \end{gathered}
\end{equation}
where $\lambda_0=k^2+\sigma(k)$ and asterisk denote the complex
conjugation. A value of $\lambda_0$ controls growth or decay of a
solution. Because $\tilde{a}(t)$ and $\tilde{b}(t)$ should be
bounded at any $k$, $\lambda_0$ does not depend on $k$. One can
easily check that~\eqref{eLinModes} also describes small
perturbation to an orbit of~\eqref{eOdeSystem}. It means that the
conditions on $\tilde{a}(t)$ and $\tilde{b}(t)$ are fulfilled when
$\lambda_0$ is equal to the largest Lyapunov exponent
of~\eqref{eOdeSystem}. Thus, we can write
\begin{equation}
  \label{eCharExp}
  \sigma(k)=\lambda_0-k^2.
\end{equation}

\begin{figure}
  \onefig{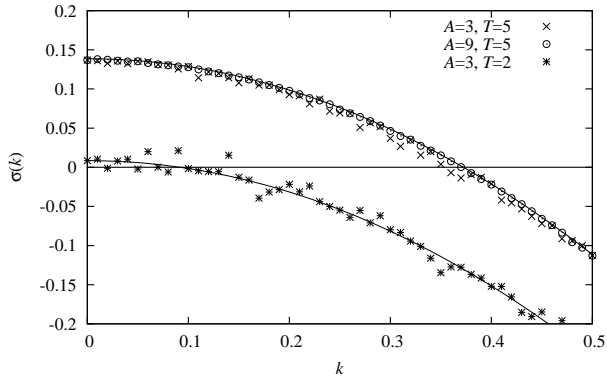}
  \caption{\label{fCharExp}Growth rate $\sigma(k)$ of
  non-uniform perturbation to a homogeneous solution of
  the system~\eqref{eTheSystem}. The solid curves are the
  graphs of~\eqref{eCharExp}: the upper one
  is $(\ln 2)/5-k^2$, the lower one is $\sigma(0)-k^2$, where
  $\sigma(0)=0.00842$. Crosses, circles and stars represent
  numerical data, computed at parameters that are shown in the
  legend.}
\end{figure}

Relation~\eqref{eCharExp} can be verified by direct numerical
computations of $\sigma(k)$. For this purpose we substitute
$\lambda_0\to\sigma(k)+k^2$ to~\eqref{eLinModes} and set there
$\sigma(k)=0$. It meas that now the amplitudes $\tilde{a}$ and
$\tilde{b}$ are allowed to grow or decay, so that the rate will be
equal to $\sigma(k)$. Given $k$, we find $\sigma(k)$ employing the
algorithm of computing of the largest Lyapunov
exponent~\cite{ParkerChua}. System~\eqref{eLinModes} is
initialized with a unit vector, and then solved together
with~\eqref{eOdeSystem} on one period $T$. After that, a norm of
the vector-solution of \eqref{eLinModes} is found and stored, and
the vector itself is normalized. When this procedure is repeated
for a sufficiently long time, the averaged logarithms of the
collected norms determine the sought $\sigma(k)$. The results are
shown in Fig.~\ref{fCharExp}. Solid lines represent theoretical
$\sigma(k)$ \eqref{eCharExp}. The upper one corresponds to a
hyperbolic chaos in \eqref{eOdeSystem} and $\lambda_0$ is found
according to~\eqref{eHypLyapunov}. The lower curve also
corresponds to chaotic oscillations of~\eqref{eOdeSystem}, that
are, however, non-hyperbolic. In this case we substitute a
computed value of $\sigma(0)$ to~\eqref{eCharExp} instead of
$\lambda_0$. Numerical data fit well the theoretical curves. As
follows from~\eqref{eCharExp} and~\eqref{eHypLyapunov},
$\sigma(k)$ does not depend on the $A$ in the regime of hyperbolic
chaos. Numerical verification confirms this.

Linear modes described by~\eqref{eLinModes} are influenced
parametrically by a chaotic force. It means that all modes with
positive $\sigma(k)$ can grow simultaneously giving rise
spatiotemporal chaos. The spectrum of lineally unstable modes with
$\sigma(k)>0$ can be found from~\eqref{eCharExp}. These modes lay
within the interval of wave numbers $0\leq k<k_\sublinmax$, where
\begin{equation}
  \label{eKmax}
  k_\sublinmax=\sqrt{\lambda_0}.
\end{equation}

If the system~\eqref{eTheSystem} is bounded by the length $L$, the
spectrum of modes allowed by the boundary
conditions~\eqref{eBoundCond} is
\begin{equation}
  \label{eEigModes}
  k_n=n\pi/L, \quad n=1,2,3,\ldots.
\end{equation}
When $L$ is small, so that $k_1 > k_\sublinmax$, there are no
unstable eigenmodes and the system demonstrates homogeneous
oscillations. Spatial structure emerges above the critical point
which can be found from the condition $k_1 = k_\sublinmax$:
$L_\subcrit=\pi/\sqrt{\lambda_0}$. Below we put attention to the
case when the local dynamics is hyperbolic. The Lyapunov exponent
$\lambda_0$ in this case is given by~\eqref{eHypLyapunov} and the
critical length reads:
\begin{equation}
  \label{eLcritHyp}
  L_\subcrit=\pi\sqrt{T/\ln 2}.
\end{equation}

\section{\label{sSptm}Spatiotemporal dynamics}

\begin{figure}
  \onefig{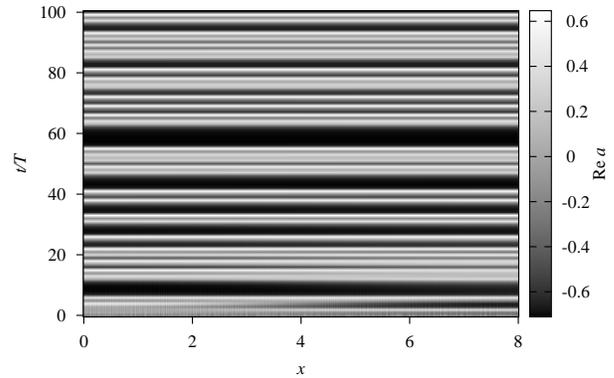}
  \caption{\label{fSptm1}Spatiotemporal dynamics
  of~\eqref{eTheSystem} at $L=8$. Grey levels
  indicate values of $\Re a$. $A=3$, $T=5$, $\epsilon=0.05$.}
\end{figure}

Let us consider some illustrations of spatiotemporal dynamics of
the system~\eqref{eTheSystem}. Figure~\ref{fSptm1} represents
homogeneous oscillations. In this and subsequent figures the space
coordinate is horizontal, time is directed vertically and grey
levels indicate values of $\Re a$ as shown by gradient bars at the
right edges of the diagrams. Layers $\Re a(x)$ are plotted at
successive steps $t_n=nT$. Critical length, according
to~\eqref{eLcritHyp}, is $L_\subcrit\approx 8.44$. The length of
the system in Fig.~\ref{fSptm1} is less then the critical value,
$L=8$. Hence, after a short transient time, it settles in a regime
of homogeneous oscillations.

\begin{figure}
  \onefig{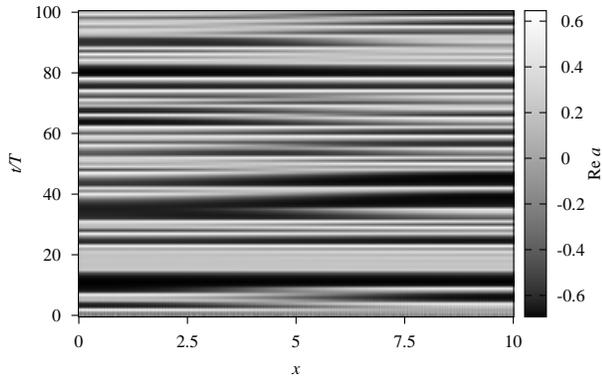}
  \caption{\label{fSptm2}Same as Fig.~\ref{fSptm2}, but at $L=10$.}
\end{figure}

In Fig.~\ref{fSptm2} the length $L=10$ is larger than
$L_\subcrit$. The first eigenmode $\cos(k_1 x)$ falls into the
instability domain and grows, destroying the homogeneity. The
first mode contains one half of the period of cosine, so if a
maximum is at the left edge of the system, a minimum appears at
the right edge and vice versa. Careful inspection of
Fig.~\ref{fSptm2} confirms this conclusion. If a horizontal
stripe, representing $\Re a(x)$ at a certain time step, is white
at the left edge, it becomes dark at the right edge.

\begin{figure}
  \onefig{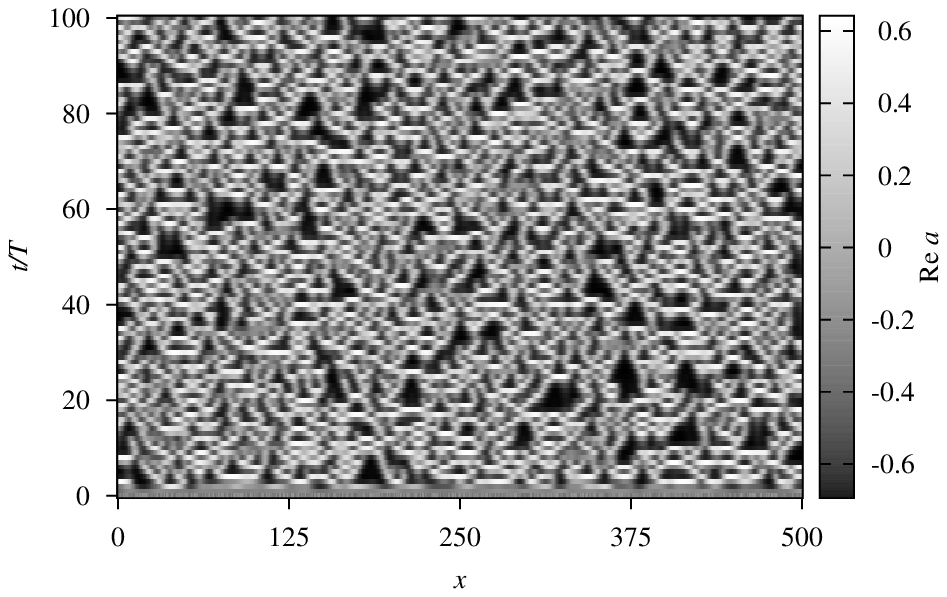}
  \caption{\label{fSptm3}$L=500$.}
\end{figure}

The result of further increase of the length up to $L=500$ is
shown in Fig.~\ref{fSptm3}. As here a lot of eigenmodes satisfy
the condition $k_n<k_\sublinmax$, they are exited and produce a
rich and complicated structure. It is interesting to note that it
reminds a structure generated by a cellular automata of Wolfram's
class 3~\cite{Wolfram86}.

\section{\label{sLyap}Lyapunov analysis}

Lyapunov exponents are average rates of expansion or contraction
in the tangent space on an attractor. They characterize the
sensitivity of motion to small perturbations; an attractor with a
positive exponent is chaotic. Also, it is important to know a
mutual orientation of expanding and contracting directions in the
tangent space at each point of the attractor. This information can
be provided by covariant Lyapunov vectors~\cite{Ginelli07}. If
there is a well defined split of the tangent space into
contracting and expanding subspaces, the dynamics is hyperbolic.
On the contrary, the dynamics is non-hyperbolic when couples of
collinear vectors from contracting and expanding subspaces can be
encountered with a non-zero probability.

In these section we compute covariant Lyapunov vectors and perform
a verification of hyperbolicity of the attractor
of~\eqref{eTheSystem}. Also we analyze Lyapunov exponents for the
system~\eqref{eTheSystem} as well as related to them Kaplan-Yorke
dimension and Kolmogorov-Sinai entropy.

\subsection{Verification of hyperbolicity at different lengths of the system}

To verify the hyperbolicity one needs to analyze expanding and
contracting directions in the tangent space on an attractor. These
directions can be found in a form of covariant Lyapunov
vectors~\cite{Ginelli07}. The method of computation of these
vectors is briefly described in Appendix.

When the covariant Lyapunov vectors are computed at some point of
the attractor, the simplest way to verify the hyperbolicity is to
estimate angles between each couple of expanding and contracting
vectors and find the smallest one. Collecting the smallest angles
for sufficiently many points, one obtains a sufficient condition
for non-hyperbolicity: the attractor is non-hyperbolic if zero
angle can be encountered with non-zero probability. But the
converse is not true. The covariant Lyapunov vectors may not be
collinear themselves, but the loss of hyperbolicity still can take
place due to a tangency of some other couple of vectors from
contracting and expanding subspaces. To take this situation into
account, a more subtle approach should be used.

Let us suppose that at some point of the attractor we have $n_\subS$
covariant Lyapunov vectors spanning the contracting tangent
subspace $\mathcal{S}$ and $n_\subU$ vectors that span the
expanding subspace $\mathcal{U}$. It is natural to assume that
$n_\subS>n_\subU$. Consider unit vectors $s\in\mathcal{S}$ and
$u\in\mathcal{U}$ and find among them a couple $s_1$ and
$u_1$ that produces the largest inner product. Arc cosine of
$s_1^\supT u_1$ is the smallest angle between subspaces, that is
denoted as $\theta_1$. Then we seek for unit vectors $s_2$ and
$u_2$ that again produce the largest inner product but with
additional requirement to be orthogonal to $s_1$ and $u_1$,
respectively. Arc cosine of their inner product is denoted as
$\theta_2$. Proceeding with this procedure, we obtain $n_\subU$
angles,
\begin{equation}
  0\leq\theta_1\leq\ldots\leq\theta_{n_u}\leq\pi/2,
\end{equation}
that are called the principal angles. Corresponding vectors $s_i$
and $u_i$ are called the principal vectors. The formal definition
of the principal angles and vectors is the
following~\cite{GolubLoan}:
\begin{equation}
  \cos \theta_k = \max_{s\in\mathcal{S}} \; \max_{u\in\mathcal{U}} \;
    s^\supT u = s^\supT_k u_k,
\end{equation}
where
\begin{equation}
  \begin{gathered}
    s^\supT s = u^\supT u = 1, \\
    s^\supT s_i = 0, \; u^\supT u_i = 0, \; i=1,\ldots,k-1.
  \end{gathered}
\end{equation}
The algorithm of computation of the principal angles is discussed
in Appendix.

Vanish of the principal angles indicate a tangency between
contracting and expanding subspaces and violation of the
hyperbolicity. A necessary and sufficient condition for the loss
of hyperbolicity is appearance of such distribution of $\theta_1$ on the
attractor that it has a non-zero value at the origin. If a system has
many degrees of freedom, several smallest principal angles can
vanish simultaneously, that means that several couples of
contracting and expanding vectors merge. A number of such angles
defines the dimension of the tangency. A necessary and
sufficient condition for the $n$-dimensional tangencies is a
non-zero probability of vanish of the sum of first $n$ principal
angles.

\begin{figure}
  \onefig{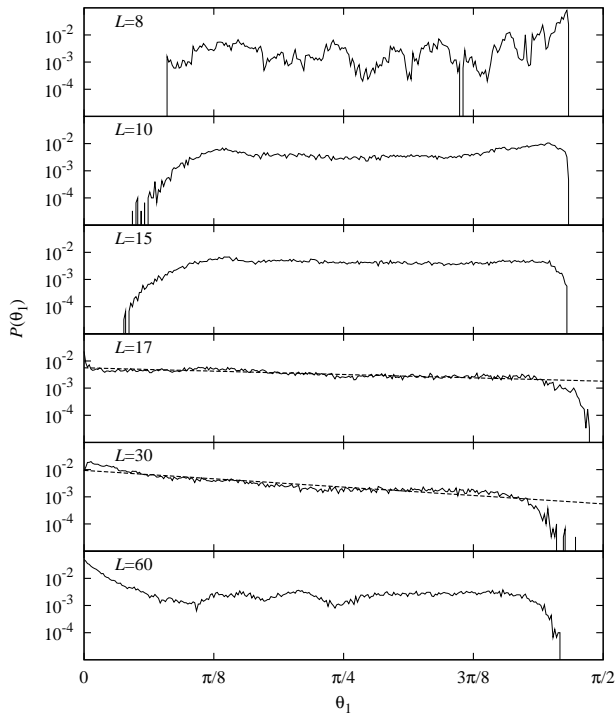}
  \caption{\label{fAngDistr}Distributions of minimal angles $\theta_1$
  between contracting
  and expanding tangent subspaces of the attractor of
  \eqref{eTheSystem}. $A=3$, $T=5$, $\epsilon=0.05$.
  The logarithmic scale is used along the ordinate axis.
  The dashed lines in the panels $L=17$, and $L=30$
  are obtained via least squares fit.
  There is one positive Lyapunov exponents at $L=8$,
  two at $L=10$ and $L=15$, three at $L=17$, five at $L=30$, and
  eleven at $L=60$. Observe the violation of hyperbolicity at $L>15$, i.e,
  when the third Lyapunov exponent becomes positive.}
\end{figure}

Figure~\ref{fAngDistr} represents distributions of $\theta_1$ for
the system~\eqref{eTheSystem} at different lengths $L$. The
equations have been solved at $\Delta t\approx 0.01$ and $\Delta x
= L/(N-1)$, where $N$ is the number of points of a numerical mesh.
$N=51$ for all $L$, except $L=60$ where $N=101$. The distributions
have been computed with resolution $300$ points. For each
distribution $10$ trajectories of the length $600T$ have been
processed with the interval $T/30$ between re-normalizations and
orthogonalizations (see Appendix for details). In the course of the
backward iterations, a time interval $500T$ is omitted as transient,
and then the angles are computed on the interval $100T$. Thus,
totaly $3000$ angles for each trajectory have been stored. The
distributions have been normalized, $\int_0^{\pi/2}P(\theta_1)=1$.

The upper curve $L=8$ in Fig.~\ref{fAngDistr} corresponds to a
spatially homogeneous case. The angles are very well localized.
Thus, the hyperbolic dynamics is observed that corresponds to the
hyperbolic dynamics of the ODE system~\eqref{eOdeSystem}. The
second curve $L=10$ represents the case of a weak inhomogeneity,
when the system is not far above the critical point $L_\subcrit$.
Observe that the distribution becomes much more smooth, compared
to the homogeneous case. It means that different configurations of
contracting and expanding subspaces are encountered with almost
equal probabilities. The distribution is still separated well from
the origin, i.e., the attractor remans hyperbolic. This is also
the case for the next distribution at $L=15$. This distribution is
even more flat than the previous one, and also it is separated
well from the origin. Notice that there are two positive Lyapunov
exponents both at $L=10$ and at $L=15$. The picture becomes
dramatically different at $L=17$, when the third Lyapunov exponent
becomes positive. The distribution occupies almost the whole range
of angles and has non-zero value at origin. The former indicates
that the attractor becomes non-hyperbolic. Moreover notice that in
the logarithmic scale the curve decays linearly from the origin.
It means that the most part of the distribution is described by an
exponential function. Similar behavior is observed at $L=30$: the
most part of the curve obeys the exponential law. The
exponents, that are equal to the slopes of the dashed approximating
lines, are $-0.72$ at $L=17$, and $-1.81$ at $L=30$, i.e., their
absolute values grow with $L$. When $L$ gets larger, as in the
panel for $L=60$, the distribution undergoes a transformation. It
acquires an extended sloping segment near the origin, while the
other part of the distribution becomes more or less flat, on average. The
attractor remains non-hyperbolic, and, moreover, the
probability to encounter the tangency of contracting and expanding
subspaces becomes larger.

\begin{figure}
  \onefig{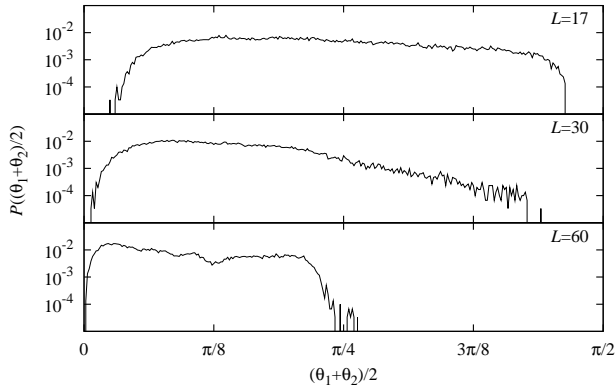}
  \caption{\label{fAngDistr2}Distributions of two principal angles $(\theta_1+\theta_2)/2$.
  Observe how curves approach the origin and touch it at
  $L=60$, that indicates the two-dimensional tangencies between
  contracting and expanding subspaces.
  The parameters are as in Fig.~\ref{fAngDistr}.}
\end{figure}

We can assume that the reorganization of the structure of
distribution, that occurs between $L=30$ and $L=60$, is associated
with emergence of the two-dimensional tangencies of contracting
and expanding subspaces. Figure~\ref{fAngDistr2} demonstrates
distributions of two first principal angles
$(\theta_1+\theta_2)/2$. The curve at $L=17$ is separated well
from the origin, so that no two-dimensional tangencies take
place. At $L=30$ the curve approaches zero much closer. Finally,
the curve at $L=60$ touches the ordinate axis confirming the
presence of the two-dimensional tangencies.

\begin{figure}
  a)\onefig{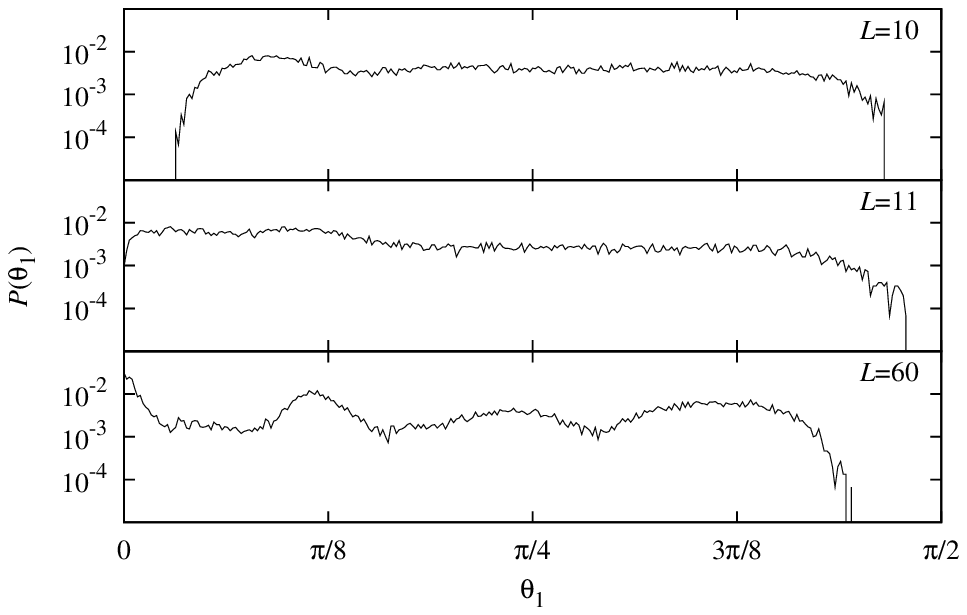}\\
  b)\onefig{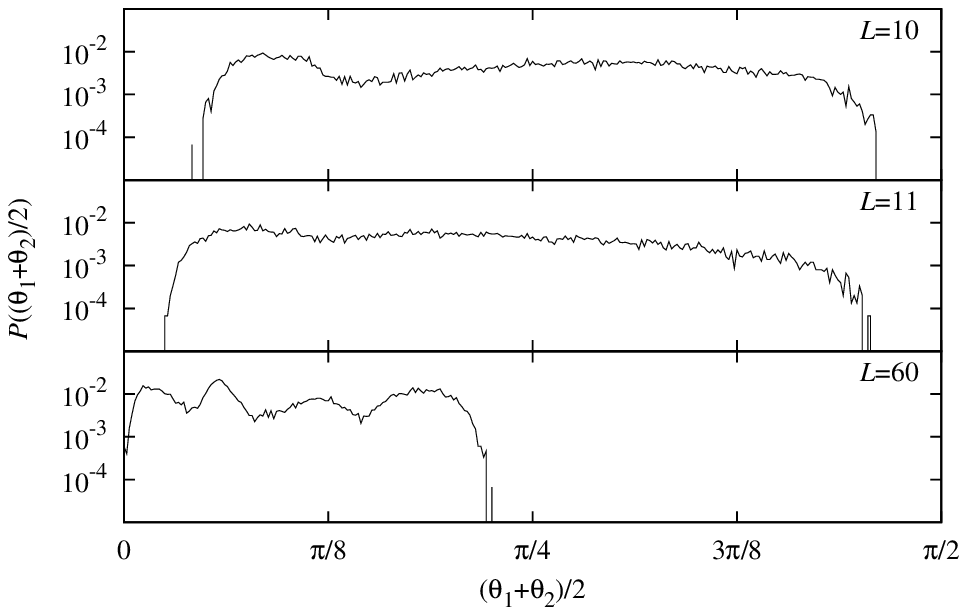}
  \caption{\label{fAngDistrT2}Distributions of $\theta_1$ and
  $(\theta_1+\theta_2)/2$, panels (a) and (b), respectively, at
  $A=8$, $T=2$, $\epsilon=0.05$. There are two positive Lyapunov exponents at $L=10$, three
  at $L=11$, and sixteen at $L=60$.
  Observe the violation of hyperbolicity
  at $L=11$ in the panel (a), and the emergence of two-dimensional tangencies
  at $L=60$ in the panel (b).}
\end{figure}

Figure~\ref{fAngDistrT2} reproduces the observed scenario at 
some other set of parameters. In panel (a) we can see that the
attractor is hyperbolic with two positive Lyapunov exponents,
curve $L=10$, while emergence of the third one results in the
violation of the hyperbolicity, curve $L=11$. Similar to the case
presented in Fig.~\ref{fAngDistr}, the distribution right above
the violation point is basically exponential, curve $L=11$, while
the further growth of $L$ results in the transformation of the
distribution, curve $L=60$. Figure~\ref{fAngDistrT2}(b) indicates
the emergence of the two-dimensional tangencies in this case: the
distributions of $(\theta_1+\theta_2)/2$ approach the origin as
$L$ grows and touch it at $L=60$.

So, we observe that the growth of $L$ first results in the
violation of hyperbolicity due to one-dimensional tangencies of
contracting and expanding subspaces, and then gives rise to
two-dimensional tangencies between these subspaces. It is natural
to suggest, that the tangencies of higher dimensions also arise
at appropriate lengthes of the system. The violation of
hyperbolicity is accompanied by the emergence of the third
positive Lyapunov exponent. Let us denote the point where the
third Lyapunov exponent passes zero as $L_2$. We suspect that the
loss of hyperbolicity takes place exactly at $L=L_2$, and below
additional evidences of this assertion are presented.

\subsection{Lyapunov exponents against the length of the system}

\begin{figure}
  \onefig{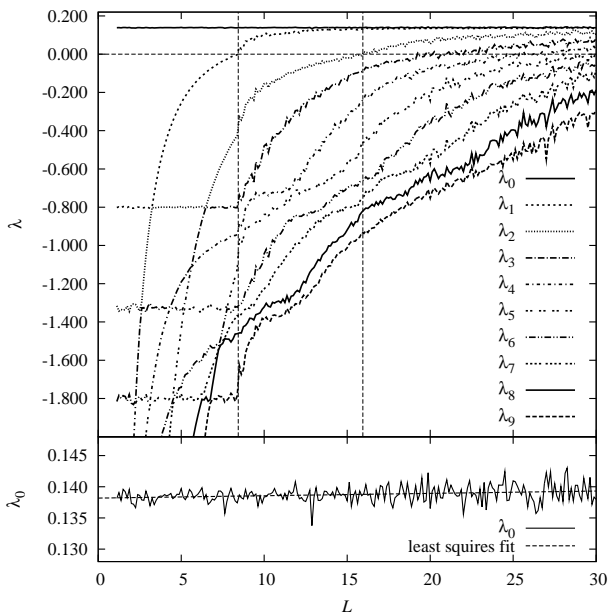}
  \caption{\label{fLyapX}Ten largest Lyapunov exponents of the
  system~\eqref{eTheSystem} against $L$. $A=3$, $T=5$,
  $\epsilon=0.05$. Vertical dashed lines mark
  $L_\subcrit\approx 8.44$ and $L_2\approx 15.9$ (the point where
  $\lambda_2=0$).
  Lower panel represents $\lambda_0$ in a large scale.
  The dashed approximating line, $4\times 10^{-5}L+0.138$,
  is obtained via least squares fit.}
\end{figure}

Figure~\ref{fLyapX} represents the Lyapunov exponents $\lambda_i$ as
functions of $L$. The plots are obtained at $N=51$ points of the
spatial mesh, $\Delta x = L/(N-1)$ and $\Delta t \approx 0.01$.
The interval between re-normalizations and orthogonalizations is
$T/30$ (see Appendix for details). Notice that the zero exponent
is absent. This is natural for the non-autonomous system we deal with.

The largest exponent $\lambda_0$ remains almost constant as $L$
varies, see the lower panel in Fig.~\ref{fLyapX}. The
approximating line, obtained via least squares fit, does not have
a noticeable slope (the slope is of the order $10^{-5}$) and is
plotted at constant value $0.138$. This is equal with a remarkable
accuracy to the theoretically predicted largest Lyapunov
exponent~\eqref{eHypLyapunov} of the corresponding ODE
system~\eqref{eOdeSystem}. When $L$ is small, the system has the
single positive exponent that corresponds to spatially homogeneous
chaotic oscillations. As $L$ grows, the second exponent
$\lambda_1$ becomes positive at $L=L_\subcrit$. This indicates the
transition to a spatially inhomogeneous solution. Further increase
of $L$ results in a cascade of passing through zero of the
exponents.

\begin{figure}
  \onefig{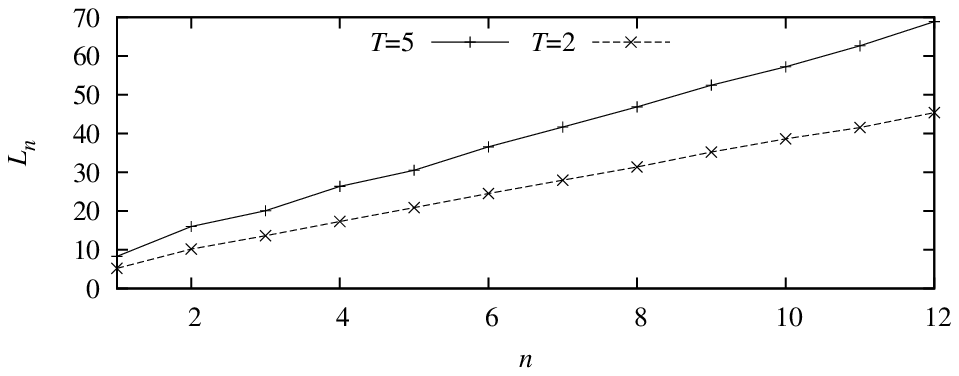}\\
  \onefig{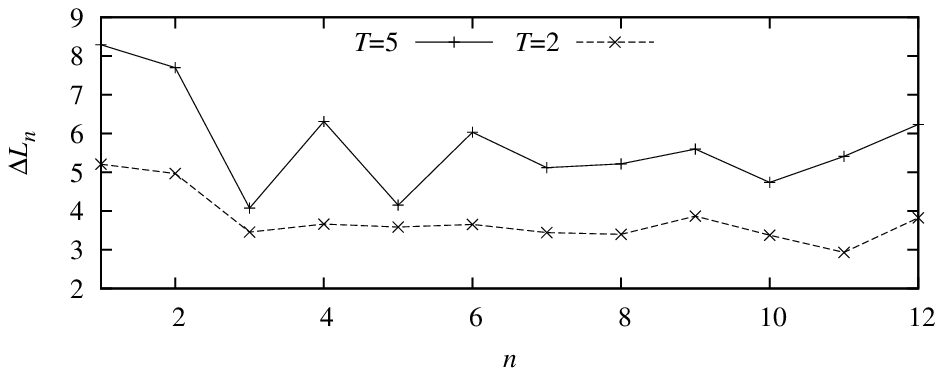}
  \caption{\label{fLz}(a) Values of length $L_n$ where corresponding
  Lyapunov exponents $\lambda_n$ pass zero,
  and (b) intervals between these points
  $\Delta L_n=L_n-L_{n-1}$
  ($\Delta L_1\equiv L_\subcrit$).
  Solid lines on both panels correspond to the parameters
  $A=3$, $T=5$, $\epsilon=0.05$, and the dashed lines represent
  parameters $A=8$, $T=2$, $\epsilon=0.05$.}
\end{figure}

Fig.~\ref{fLz}(a) shows lengthes $L_n$ where corresponding
Lyapunov exponents $\lambda_n$ vanish. Two lines correspond to two
sets of parameters of the system. One can see that $L_n$ depends
linearly on $n$. It means that the number of positive exponents
also linearly, on average, grows with $L$. In Fig.~\ref{fLz}(b)
the intervals $\Delta L_n=L_n-L_{n-1}$ are plotted ($\Delta
L_1\equiv L_\subcrit$). Notice that $\Delta L_2\approx \Delta L_1$
and these two values are lager then the others $\Delta L_n$. We
attribute this to the transition to a non-hyperbolic attractor
that takes place at $L_2$.

\subsection{Kaplan-Yorke dimension and Kolmogorov-Sinai entropy}

\begin{figure}
  a)\onefig{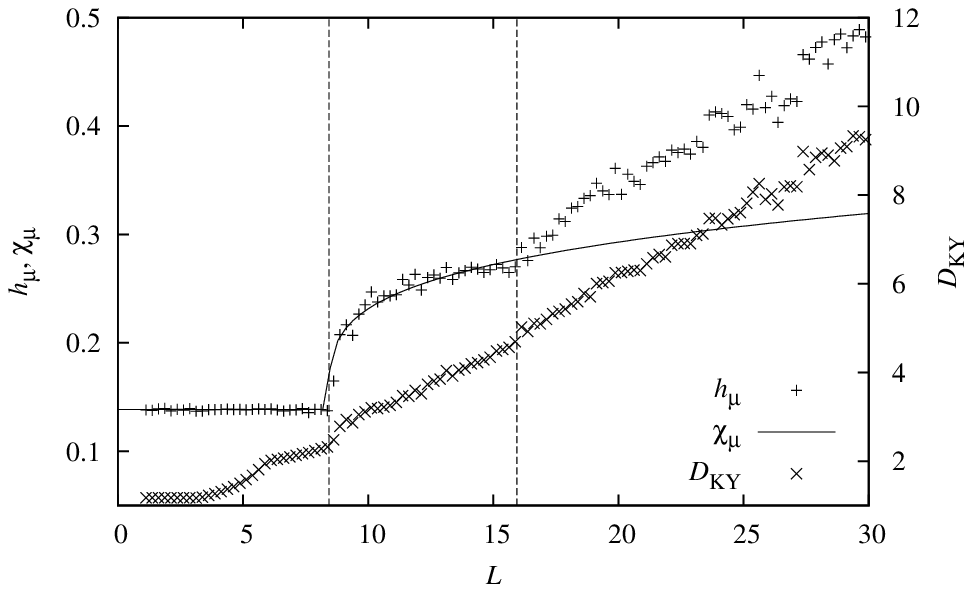}\\
  b)\onefig{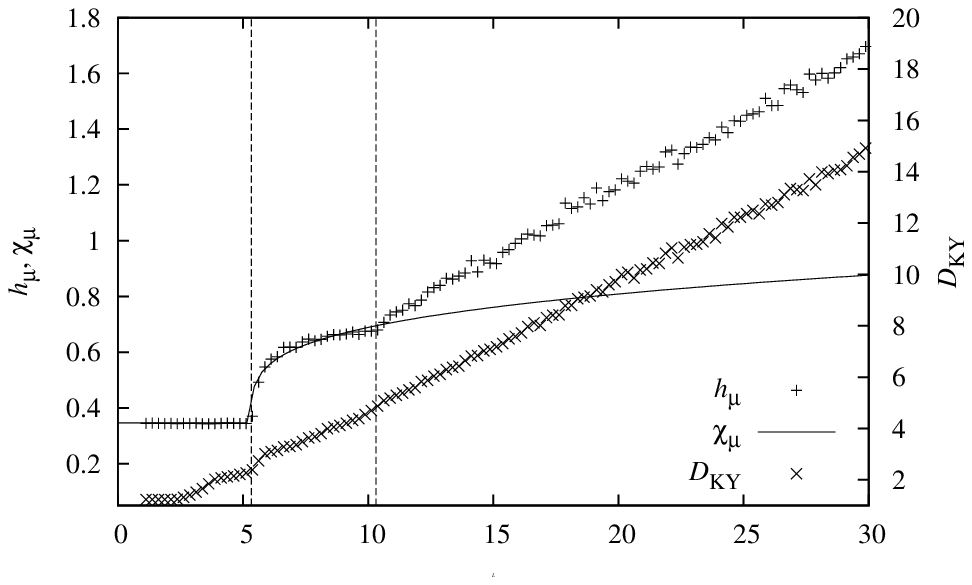}
  \caption{\label{fKYDimX}Kaplan-Yorke dimension $D_\subkapyor$,
  upper estimate $h_\mu$ for the Kolmogorov-Sinai entropy, and its
  power low approximation $\chi_\mu$~\eqref{eEntropy} against $L$.
  (a) $A=3$, $T=5$,
  $\epsilon=0.05$. (b)
  $A=8$, $T=2$,
  $\epsilon=0.05$.
  Vertical dashed lines mark
  $L_\subcrit$ and~$L_2$.}
\end{figure}

Figure~\ref{fKYDimX} illustrates the Kaplan-Yorke or Lyapunov
dimension $D_\subkapyor$~\cite{Schuster84,Ott93}, and the sum of
positive Lyapunov exponents $h_\mu$, which is an upper estimate
for the Kolmogorov-Sinai or metric
entropy~\cite{Schuster84,Ott93}. Two panels are obtained for
different sets of parameters. Vertical dashed lines mark the point
$L_\subcrit$ of transition to the spatially inhomogeneous
attractor, and the point $L_2$, where the third Lyapunov exponent
passes zero so that the attractor becomes non-hyperbolic.

Let us consider $h_\mu$ in more details. It is known that for a
hyperbolic attractor $h_\mu$ is equal to its Kolmogorov-Sinai
entropy, while for a generic chaotic attractor this is an upper
estimate for the entropy~\cite{Ott93}. Because our system is
hyperbolic at $L<L_2$, we can use $h_\mu$ to construct a function
which approximates the entropy at least on this interval. Below
$L_\subcrit$ we have $h_\mu=\lambda_0$, while above this point $h_\mu$
demonstrates a power law behavior. Thus, employing the least
squares fit, we obtain a function, approximating $h_\mu$ as
\begin{equation}
  \label{eEntropy}
  \chi_\mu(L)=\left\{
    \begin{array}{ll}
      \lambda_0 & L\leq L_\subcrit,  \\
      \alpha(L-L_\subcrit)^\gamma+\lambda_0 & L>L_\subcrit,
    \end{array}
    \right.
\end{equation}
where $\alpha=0.083$ and $\gamma=0.25$ for Fig.~\ref{fKYDimX}(a)
and $\alpha=0.229$ and $\gamma=0.26$ for Fig.~\ref{fKYDimX}(b).
The indices $\gamma$ computed for different parameter sets are,
perhaps, identical (small difference can be attributed to errors
of computations).

The power law approximation~\eqref{eEntropy} agrees very well with
the numerical curve $h_\mu$ at $L<L_2$, and at $L=L_2$ a
bifurcation occurs that is associated with the loss of
hyperbolicity. There are two possibilities above this point. The
first one is that the Eq.~\eqref{eEntropy} still gives correct value
of the entropy, while $h_\mu$ serves as an upper estimate. The
other possibility is that $h_\mu$ correctly represents the
entropy, while the approximation~\eqref{eEntropy} becomes
inappropriate. Anyway, both of these variants fit well with our
conclusion that the system loses the hyperbolicity at $L=L_2$.

Above $L_2$ the entropy $h_\mu$ grows linearly with the length, 
as well as the dimension.
A number of positive Lyapunov exponents also 
demonstrates a linear growth as follows from the linear growth of $L_n$
in Fig.~\ref{fLz}(a). This is a typical phenomenon for extensive fully
developed chaos in extended systems. In particular, the linear
growth of $D_\subkapyor(L)$ and $h_\mu(L)$ was reported for
coupled map lattices~\cite{Kaneko89}, for Kuramoto-Sivashinsky
(KS) equation~\cite{Mann85} and for complex Ginzburg-Landau (CGL)
equation~\cite{Keefe89}. Also, the linear growth of $D_\subkapyor$
was demonstrated for a chaotic attractor of coupled
Ginzburg-Landau equations~\cite{JungParl99}. It can be explained
by exponential decay of spatial correlations. Two points with
space separation larger
than the correlation length, move independently, so that
the system can be roughly represented by a direct product of
independent subsystems~\cite{Kaneko89}. Thus, the additivity is
observed: the growth of $L$ merely results in the proportional
increase of the characteristic values.

% a) Slope of dim 0.32502, Slope of ent 0.0145281
% b) Slope of dim 0.505775, Slope of ent 0.0499488

\subsection{Spectra of Lyapunov exponents}

\begin{figure}
  a)\onefig{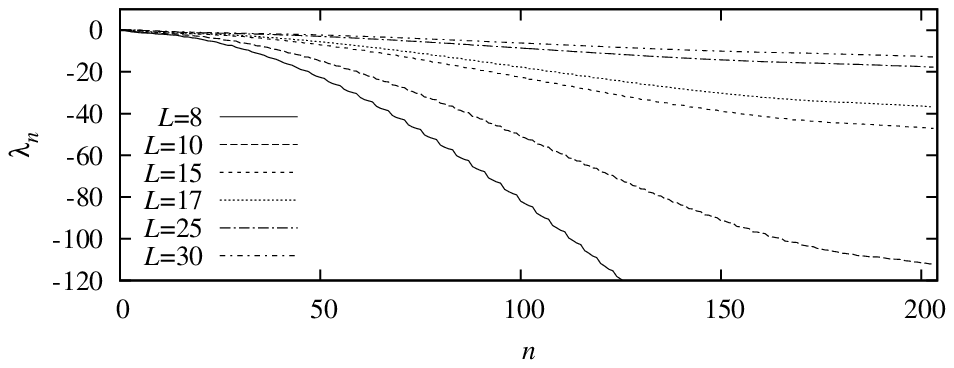}\\
  b)\onefig{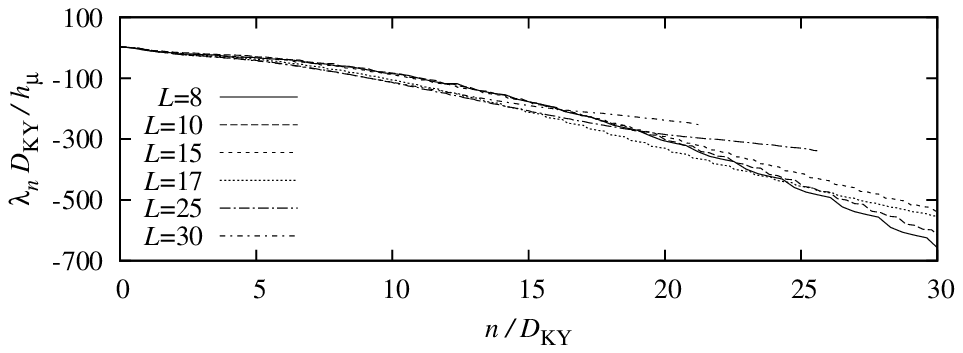}\\
  c)\onefig{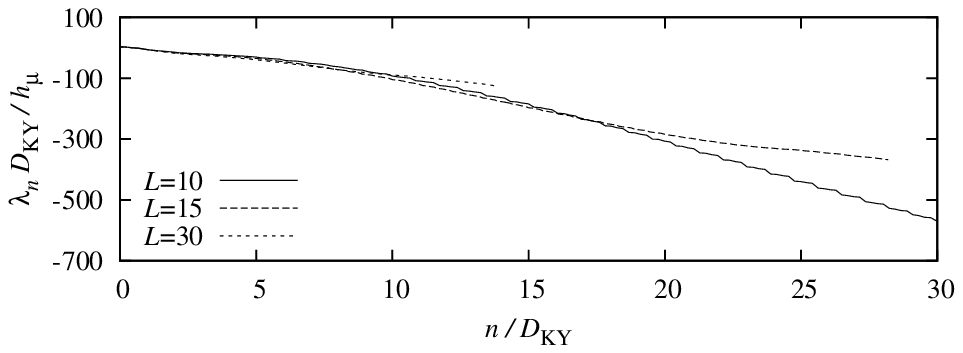}
  \caption{\label{fLyapSpectr}(a) Spectra of Lyapunov exponents, and
  (b) scaled spectra at $A=3$, $T=5$,
  $\epsilon=0.05$. (c) Scaled spectra at $A=8$, $T=2$,
  $\epsilon=0.05$.}
\end{figure}

Figure~\ref{fLyapSpectr}(a) demonstrates spectra of Lyapunov
exponents at different $L$. The first curve $L=8$ corresponds to a
spatially homogeneous case when oscillations in all spatial points
are synchronized and can be described by~\eqref{eOdeSystem}. There
is one positive Lyapunov exponent. As one can see from the figure,
the minor negative exponents have vary large absolute values. It
means that only a few spatial modes are actually involved in the
dynamics, while the most of modes are highly damped. When $L$
grows, more Lyapunov exponents becomes positive and the remaining
negative exponents approach the axis of abscissas so that their
absolute values become smaller. In the other words, more spatial
modes participate in the dynamics. The separation of modes
involved and not involved in the observable dynamics is studied in
Ref.~\cite{HongTakGin08}. For a dissipative chaotic system is
shown to exists a splitting of the tangent space into physical
modes, responsible for the observable dynamics, and hyperbolically
isolated from them highly damped non-physical modes that do not
bring an essential information about the dynamics.

For a fully developed spatiotemporal chaos a Lyapunov spectrum
scaled as $(D_\subkapyor / h_\mu)\lambda(n/D_\subkapyor)$ is known
to tend to a limiting curve at $L\to\infty$. In particular, this
was reported for coupled map lattice~\cite{Kaneko89}, for
Kuramoto-Sivashinsky (KS) equation~\cite{Mann85} and for complex
Ginzburg-Landau (CGL) equation~\cite{Keefe89}.
Figures~\ref{fLyapSpectr}(b) and (c) represent the verification of
this property for the system~\eqref{eTheSystem} at two sets of
parameters. One can see high correspondence of curves, obtained at
different~$L$.

\section{\label{sRes}Summary and conclusion}

We considered an extended system whose local dynamics is
hyperbolic and spatial coupling is introduced via diffusion. A
numerical verification of hyperbolicity of the attractor of this
system was performed. The test was based on the computation of
distributions of principal angles between contracting and
expanding tangent subspaces of the attractor. The analysis
revealed that the hyperbolicity is inherent only to a
low-dimensional chaos observed at sufficiently small lengths of
the system.

The dynamics is obviously hyperbolic when oscillations are
homogeneous in space, because each spatial cell merely reproduces
the oscillations of a partial ODE system that is known to be
hyperbolic. This regime is characterized by a single positive
Lyapunov exponent. The hyperbolicity survives when the length gets
larger, so that the first spatial mode allowed by boundary
conditions becomes linearly unstable, and the oscillations becomes
inhomogeneous. This transition is accompanied by the emergence of
the second positive Lyapunov exponent. Further growth of the
length results in the emergence of the third positive Lyapunov
exponent. In this point the violation of hyperbolicity takes
place.

Beyond the point of the hyperbolicity loss, the system
demonstrates an extensive spatiotemporal chaos that is
characterized by a fast decay of a spatial correlation. We
verified several standard criteria and observed behavior that is
typical for many others extended chaotic systems. Namely, the
number of positive Lyapunov exponents, the sum of positive
exponents (this value is an upper estimate for Kolmogorov-Sinai
entropy), and the Kaplan-Yorke dimension grow linearly against the
length of the system. Spectrum of the Lyapunov exponents, being
properly rescaled, tends to a limiting curve as the length grows.

So, if the length of the system grows and the third Lyapunov
exponent becomes positive, we register the violation of
hyperbolicity due to the emergence of one-dimensional
intersections of contracting and expanding tangent subspaces of
the attractor. If the length continues to increase, along with
one-dimensional intersections, we observe two-dimensional ones.
This is a stronger type of the hyperbolicity violation, because
there is higher probability for the perturbation to be transferred
between contracting and expanding subspaces. We expect that the
intersections of higher dimensions also take place as the length
diverges. It is interesting to study the violation of
hyperbolicity in the thermodynamic limit. If the number of modes
involved in the dynamics is infinite, the maximal dimension of the
intersections may be infinite too or it can have a finite value.
The first case can be termed as a strong violation, because the
capacity of set of the merging vectors from contracting and
expanding subspaces is comparable with the capacity of the whole
set of degrees of freedom. Hence, the probability for the
perturbation to be transferred between contracting and expanding
subspaces is non-zero. The second case can be termed as a weak
violation. Though the intersections take place, the number of
merging directions per degree of freedom is zero. Thus, the
probability of the perturbation transfer vanishes.

\begin{acknowledgments}
PVK acknowledges support from RFBR-DFG grant
No~08-02-91963, and SPK acknowledges support from RFBR grant
No~09-02-00426.
\end{acknowledgments}

\appendix*

\section{\label{sAppx}Computation of Lyapunov exponents, covariant Lyapunov vectors
and angles between subspaces}

To compute Lyapunov exponents, we apply an algorithm based on the
QR decomposition. See
Refs.~\cite{EckRuell85,GeistParlLaut90,Skokos08} for the details
of the algorithm, and Ref.~\cite{GolubLoan} for an idea of the QR
decomposition.

First of all, equations for small perturbations $\tilde{a}(x,t)$
and $\tilde{b}(x,t)$ to a trajectory $a(x,t)$ and $b(x,t)$
of~\eqref{eTheSystem} are required:
\begin{equation}
  \label{eLinTheSystem}
  \begin{gathered}
    \partialt \tilde{a} = A \cos (2\pi t/T) \tilde{a} -
      2|a|^2 \tilde{a} - a^2 \tilde{a}^* - \myii \epsilon \tilde{b} + \partialxx \tilde{a}, \\
    \partialt \tilde{b} = -A \cos (2\pi t/T) \tilde{b}  -
      2|b|^2 \tilde{b} - b^2 \tilde{b}^* - 2\myii \epsilon a\tilde{a} + \partialxx \tilde{b},
  \end{gathered}
\end{equation}
where asterisk denotes the complex conjugation. To compute
$M_\lambda$ Lyapunov exponents, we need $M_\lambda$ exemplars of
the linear equation sets~\eqref{eLinTheSystem}, which are
initialized by an orthogonal set of random unit vectors of the
length $4N$, where $N$ is the number of points of a numerical
mesh. Basic system~\eqref{eTheSystem} is also initialized and
advanced along a trajectory for a sufficiently long time to arrive
at the attractor. Then the basic system is solved simultaneously
with $M_\lambda$ linear equation sets during some time interval.
The more Lyapunov exponents are required, the shorter interval
should be taken, because minor negative Lyapunov exponents can
have vary large absolute values so that the corresponding
solutions of linear subsystems decay very fast. $M_\lambda$
resulting vectors are then considered as columns of a matrix that
is decomposed into an orthogonal matrix $\mtr{Q}$ and an upper
triangular matrix $\mtr{R}$. (An algorithm based on the
Householder rotation is used~\cite{GolubLoan}.) Logarithms of
$M_\lambda$ diagonal elements of the $\mtr{R}$ are collected,
while $M_\lambda$ columns of the $\mtr{Q}$ are used to
re-initialize linear systems. Then this procedure is repeated.
Averaged logarithms of diagonal elements of $\mtr{R}$ converge to
Lyapunov exponents.

To compute covariant Lyapunov vectors according to the method
recently reported in Ref.~\cite{Ginelli07}, we must do the similar
things. After initialization of the equations, we make several
steps $n_0$ accompanied by the QR procedure, but without storing
elements of $\mtr{R}$, to obtain a good matrix $\mtr{Q}_{n_0}$.
``A good'' means that each linear subspace $\mathcal{S}_{n_0}^j$,
$j=1,2,\ldots 4N$, spanned by first $j$ vector-columns of
$\mtr{Q}_{n_0}$, contains $j$-th expanding (or contracting)
direction of the tangent space at $n_0$. Starting from $n_0$, we
make some more steps and arrive at $n_1$. Here we have a matrix
$\mtr{Q}_{n_1}$ with columns that determine subspaces
$\mathcal{S}_{n_1}^j$. Our aim now is to define arbitrary unit
vectors belonging to these subspaces, $\vec{u}_{n_1}^j\in
\mathcal{S}_{n_1}^j$, $j=1,2\ldots 4N$. In fact, we just need to
generate a random upper triangular matrix $\mtr{C}_{n_1}$, whose
size coincides with $\mtr{R}$, and columns are normalized by 1.
$j$-th column of $\mtr{C}_{n_1}$ contains coordinates of
$\vec{u}_{n_1}^j$ with respect to the basis $\mtr{Q}_{n_1}$. In
the other words
\begin{equation}
  \label{eUQC}
  \mtr{U}_n = \mtr{Q}_n \mtr{C}_n,
\end{equation}
where $\mtr{U}_n=\{\vec{u}_n^1,
\vec{u}_n^2,\ldots,\vec{u}_n^{4N}\}$. Starting from
$\mtr{C}_{n_1}$, we perform backward iterations
$\mtr{C}_{n-1}=\mtr{R}^{-1}_n \mtr{C}_n$ accompanied by
re-normalization of columns of $\mtr{C}_n$. Collecting and
averaging the negative logarithms of the norms, we obtain Lyapunov
exponents. Under these iterations the vectors $\vec{u}_n^j$,
represented by columns of the $\mtr{C}_n$, are aligned with the
most expanding directions of subspaces $\mathcal{S}_n^j$. These
directions are associated with corresponding Lyapunov exponents.
Because we go back in time, the highest Lyapunov exponents do not
dominate this alignment. If the number of steps from $n_1$ to
$n_0$ is sufficiently large, getting back at $n_0$, we obtain the
matrix $\mtr{C}_{n_0}$ with coordinates of covariant Lyapunov
vectors $\mtr{U}_{n_0}$, pointing expanding and contracting
directions of the tangent space at $n_0$. Explicit form of
$\mtr{U}_{n_0}$ can be found from~\eqref{eUQC}. Computed in
parallel, the Lyapunov exponents allow to distinguish expanding
and contracting directions.

In practice, computing the covariant Lyapunov vectors for a system
of many degrees of freedom, we must deal with very large arrays of
data. For the backward procedure to be performed, $m=n_1-n_0$
matrices $\mtr{R}$ should be stored. The time interval between
successive QR decompositions should be sufficiently small to treat
minor Lyapunov exponents and corresponding vectors accurately,
while the duration of the backward procedure must be long because
the vectors are found to converge sufficiently slow. As a result,
an array of  matrices $\mtr{R}$ runs up to several gigabytes. We
recall that on 32-bit platforms the physical limit of an
addressable memory is 4Gb, while the memory actually available for
programs is even less. It means that we can not store such array
in memory and need to write it to a file. (Otherwise, one can
employ a 64-bit platform with appropriate amount of memory, of
course.) Moreover, the file must be written in a binary format.
The usual text format is not a saving so that an extremely large
file can be obtained.

According to Eq.~\eqref{eUQC}, we need $\mtr{Q}_{n_0}$ to restore
covariant Lyapunov vectors in the original phase space. It meas
that an array of $m$ matrices $\mtr{Q}$ must also be stored.
Hopefully, this is not needed. The transformation~\eqref{eUQC}
preserves angles because matrices $\mtr{Q}_n$ are orthogonal.
Thus, we do not need the $\mtr{U}_n$ to analyze the structure of
the tangent space. Identical information about this space can be
extracted directly from the column-space of $\mtr{C}_n$.

To compute the $\mtr{C}_n$ we apply a two-pass procedure. First,
we solve the equations and perform QR decompositions during a
sufficiently long time, saving obtained matrices $\mtr{R}_n$ to a
file. Then, on the second pass, we generate random matrix
$\mtr{C}_{n_1}$, see the details above, and perform the backward
iterations, reading $\mtr{R}_n$ from the file from the end to the
beginning. When a sufficiently large number of transient
iterations are made, we start to compute angles between
contracting and expanding subspaces of the column-space of
$\mtr{C}_n$ until arrive at the beginning of the file of
$\mtr{R}_n$.

The algorithm of computation of the angles between subspaces, so
called principal angles, can be found, e.g., in
Refs.~\cite{GolubLoan,KnyArg02}. Consider a matrix $\mtr{C}_n$.
First of all, its columns must be classified as vectors associated
with contracting and expanding directions of the tangent space,
according to signs of corresponding Lyapunov exponents. Thus we
obtain a matrix $\mtr{S}$ comprising of $n_s$ covariant Lyapunov
vectors from the contracting subspace and a matrix $\mtr{U}$ that
consists of $n_u$ vectors of the expanding subspace. It is
naturally to assume that $n_s>n_u$. For both of these matrices we
compute the QR factorizations $\mtr{S}=\mtr{Q}_s \mtr{R}_s$,
$\mtr{U}=\mtr{Q}_u \mtr{R}_u$, and then compose the
matrix~$\mtr{M}$:
\begin{equation}
  \label{eMMatrix}
  \mtr{M}=\mtr{Q}_s^\supT \mtr{Q}_u.
\end{equation}
Cosines of the sought principal angles $\theta_i$,
($i=1,\ldots,n_u$) are equal to the singular values of the
$\mtr{M}$, that can be easily computed, see
e.g.~\cite{GolubLoan,Recipes}.

This algorithm is known to fail to accurately compute very small
angles, and in Ref.~\cite{KnyArg02} an improved version is
suggested. But, nevertheless, we use the standard algorithm,
because the extremely high accuracy is not needed for our
purposes.

\bibliography{hyperspace}

\end{document}